\documentclass[acmsmall]{acmart}
\renewcommand\footnotetextcopyrightpermission[1]{} 
\AtBeginDocument{%
  }

\usepackage{xcite}
\usepackage{soul}
\usepackage{url}

\usepackage[utf8]{inputenc}
\usepackage{tabularx,ragged2e,booktabs,caption}

\usepackage{subfigure}
\usepackage{amsmath}
\usepackage{booktabs}
\usepackage{algorithm}
\usepackage{algorithmic}
\usepackage{graphicx}
\usepackage{pifont}
\usepackage{color,soul}
\usepackage{xcolor}
\usepackage{setspace,lipsum}

\usepackage{tikz}
\makeatletter
\g@addto@macro\normalsize{%
  \setlength\abovedisplayskip{6pt}
  \setlength\belowdisplayskip{6pt}
  \setlength\abovedisplayshortskip{6pt}
  \setlength\belowdisplayshortskip{6pt}
}
\makeatother

\usepackage{comment}
\usepackage{diagbox}

\usepackage{array}
\usepackage{rotating}
\definecolor{myellow}{RGB}{207, 125, 3}

\usepackage{adjustbox}
\usepackage{multirow}
\usepackage{colortbl}
\usepackage{amssymb}
\usepackage{color,soul}
\usepackage{threeparttablex}
\usepackage{listings}
\usepackage{wasysym}
\usepackage{pifont}
\usepackage{float}
\usepackage[symbol]{footmisc}

\definecolor{LightGray}{gray}{0.92}

\usepackage{threeparttable}
\usepackage{booktabs,tabulary}
\usepackage{siunitx}

\definecolor{dkgreen}{rgb}{0,0.6,0}
\definecolor{gray}{rgb}{0.5,0.5,0.5}
\definecolor{mauve}{rgb}{0.58,0,0.82}

\usepackage{tcolorbox}

\usepackage{fancyhdr}
\begin{document}


\title{SecBench: A Comprehensive Multi-Dimensional Benchmarking Dataset for LLMs in Cybersecurity}


\author{Pengfei Jing}
\affiliation{%
 \institution{The Hong Kong Polytechnic University, Tencent Security Keen Lab}
 \country{China}
}

\author{Mengyun Tang}
\affiliation{%
 \institution{Tencent Zhuque Lab}
 \country{China}
}

\author{Xiaorong Shi}
\affiliation{%
 \institution{Tencent Zhuque Lab}
 \country{China}
}

\author{Xing Zheng}
\affiliation{%
 \institution{Tencent Zhuque Lab}
 \country{China}
}

\author{Sen Nie}
\affiliation{%
 \institution{Tencent Security Keen Lab}
 \country{China}
}

\author{Shi Wu}
\affiliation{%
 \institution{Tencent Security Keen Lab}
 \country{China}
}

\author{Yong Yang}
\affiliation{%
 \institution{Tencent Security Platform and Department}
 \country{China}
}

\author{Xiapu Luo}
\affiliation{%
 \institution{The Hong Kong Polytechnic University}
 \country{China}
}

%


\begin{abstract}


Evaluating Large Language Models (LLMs) is crucial for understanding their capabilities and limitations across various applications, including natural language processing and code generation. Existing benchmarks like MMLU, C-Eval, and HumanEval assess general LLM performance but lack focus on specific expert domains such as cybersecurity. Previous attempts to create cybersecurity datasets have faced limitations, including insufficient data volume and a reliance on multiple-choice questions (MCQs). To address these gaps, we propose SecBench, a multi-dimensional benchmarking dataset designed to evaluate LLMs in the cybersecurity domain. SecBench includes questions in various formats (MCQs and short-answer questions (SAQs)), at different capability levels (Knowledge Retention and Logical Reasoning), in multiple languages (Chinese and English), and across various sub-domains. The dataset was constructed by collecting high-quality data from open sources and organizing a Cybersecurity Question Design Contest, resulting in 44,823 MCQs and 3,087 SAQs. Particularly, we used the powerful while cost-effective LLMs to (1). label the data and (2). constructing a grading agent for automatic evaluation of SAQs. Benchmarking results on 16 SOTA LLMs demonstrate the usability of SecBench, which is arguably the largest and most comprehensive benchmark dataset for LLMs in cybersecurity. More information about SecBench can be found at our website \cite{secbench_site}, and the dataset can be accessed via the artifact link \cite{SecBench_artifact}.

\end{abstract}

\maketitle


\section{Introduction}
\label{sec:intro}

Evaluating Large Language Models (LLMs) is essential for understanding their capabilities and limitations, as these models play a significant role in various applications, from natural language processing to code generation. The importance of evaluating LLMs lies in ensuring their reliable and effective performance across diverse tasks while identifying areas for improvement.
Many benchmarks have been developed to assess different aspects of LLM performance, such as the MMLU benchmark for general knowledge and reasoning \cite{mmlu}, C-Eval for the Chinese context \cite{ceval}, and HumanEval for code generation and completion \cite{humaneval}. 
These benchmarks collectively provide a comprehensive framework for evaluating the multifaceted capabilities of LLMs. However, while these benchmarks focus on general capabilities, it is also crucial to evaluate LLMs in specific expert domains, such as cybersecurity.
Previous studies have attempted to establish datasets for this purpose \cite{bhusal2024secure_Benchmark_SECURE, liu2023secqa_Benchmark_SecQA, tihanyi2024_CyberMetric, liu2024_CyberBench}, but they face limitations, including insufficient evaluation data volume and a predominant use of multiple-choice questions (MCQs). A more challenging task, the short-answer question (SAQ), which requires the model to generate its own answer, has not been explored in these studies.

Towards constructing a more comprehensive dataset and benchmarking large language models (LLMs) in cybersecurity, we propose SecBench, a multi-dimensional benchmarking dataset designed to evaluate LLMs in the cybersecurity domain.
Specifically, SecBench assesses LLMs with questions in various formats (multiple-choice questions (MCQs) and short-answer questions (SAQs)), at different capability levels (Knowledge Retention (KR) and Logical Reasoning (LR)), in multiple languages (Chinese and English), and across various sub-domains, thereby covering a wide range of knowledge in cybersecurity.
To construct such an extensive dataset, we began by collecting high-quality data from open sources, resulting in an initial dataset of 10,551 MCQs.
To further expand this dataset both qualitatively and quantitatively, we organized a \emph{Cybersecurity Question Design Contest} aimed at collecting high-quality questions from the public.
After filtering and processing the data collected from the contest, we obtained an additional 34,272 MCQs and 3,087 SAQs.
Additionally, we utilized a powerful LLM, GPT-4, to automatically label the collected data according to their most relevant capability level and domain.
Following the labeling process, we derived SecBench, which consists of 44,823 MCQs and 3,087 SAQs, each well-labeled with detailed metadata.

To achieve automatic evaluation of SAQs, we employed another powerful yet cost-efficient LLM, GPT-4o-mini, as a \emph{grading agent} to automatically grade the tested LLMs' answers based on the question stem and ground truth (correct answer) provided by SecBench.
For evaluation, we conducted benchmarking of 16 State-of-the-Art LLMs on SecBench, demonstrating the usability of SecBench both qualitatively and quantitatively.

The remainder of this paper is structured as follows.
\S\ref{sec:background} provides the necessary background information.
We introduce the design of SecBench in \S\ref{sec:approach} and detail the dataset construction process in \S\ref{sec:dataset_process}.
Then we present the benchmarking results in \S\ref{sec:evaluation}, discussion in \S\ref{sec:discussion}, and conclude with \S\ref{sec:conclusion}.
%
More information about SecBench can be found at our website \cite{secbench_site}, and the dataset can be accessed via the artifact link \cite{SecBench_artifact}.

\section{Background}
\label{sec:background}

\textbf{Benchmarking LLMs.} Evaluating Large Language Models (LLMs) is crucial for understanding their capabilities and limitations, as these models have become increasingly influential in various applications, from natural language processing to code generation and beyond. The significance of evaluating LLMs lies in ensuring that they perform reliably and effectively in diverse tasks, while also identifying areas for improvement. Several popular benchmarks have been developed to assess different aspects of LLM performance. For instance, the MMLU benchmark evaluates general knowledge and reasoning across a wide range of subjects \cite{mmlu}. C-Eval \cite{ceval} focuses on LLM's capability in the specific Chinese context. HumanEval \cite{humaneval} is designed to assess code generation and completion tasks.
These benchmarks collectively provide a comprehensive framework for evaluating the multifaceted capabilities of LLMs.

\textbf{Benchmarking LLM in Cybersecurity.} The benchmarks discussed earlier primarily focus on assessing the \emph{general capabilities} of LLMs. However, since LLMs are often fine-tuned for specific expert domains, it is crucial to evaluate their performance across various specialized fields. In the context of cybersecurity, previous studies have attempted to establish datasets to assess LLMs' knowledge in this particular domain \cite{bhusal2024secure_Benchmark_SECURE, liu2023secqa_Benchmark_SecQA, tihanyi2024_CyberMetric, liu2024_CyberBench}. Unfortunately, these studies face two main limitations. First, the average volume of evaluation data is only at the thousand-level, which may not be sufficient to provide a comprehensive benchmark. Second, the question design in previous works predominantly follows the multiple-choice question (MCQ) format, which merely requires the model to select the correct answer from given options. 
However, a more challenging task, the short-answer question (SAQ), which requires the model to generate its own answer rather than choosing from existing ones, has not been explored in previous studies.

\section{SecBench Design}
\label{sec:approach}

\begin{figure}[t]
\centering
\includegraphics[width = 0.99\linewidth]{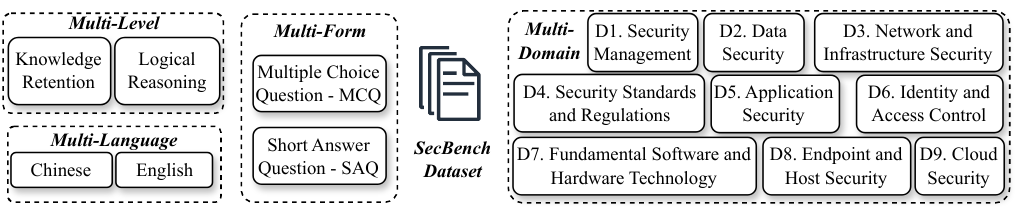}
\centering
\vspace{-2ex}
\caption{SecBench: A multi-level, multi-language, multi-form, and multi-domain benchmarking dataset for LLM in Cybersecurity.}
\label{fig:secbench_dimensions}
\vspace{-1.6ex}
\end{figure}

Fig.\ref{fig:secbench_dimensions} shows the overview of the SecBench design: it is a comprehensive benchmarking dataset aiming to benchmark LLM's capability in cybersecurity from \emph{Multi-Level}, \emph{Multi-Language}, \emph{Multi-Form}, \emph{Multi-Domain}.

\textbf{Multi-Level.}
We devise the capability of LLM in cybersecurity into two different levels: \emph{Knowledge Retention - KR} and \emph{Logical Reasoning - LR}.
Among the two, knowledge retention examines the LLM's ability to retain existing knowledge. The content of such questions is relatively straightforward and does not involve complex reasoning.
On the other hand, logical reasoning assesses the LLM's ability to infer the correct answer based on the given information. The difficulty of these questions is relatively higher and better demonstrates the model's capability to handle complex problems.

\textbf{Multi-Language.}
SecBench includes questions of two mainstream languages - Chinese and English, to present a more comprehensive benchmark.

\textbf{Multi-Form.}
Unlike previous works that constructed only multiple-choice questions (MCQs) \cite{bhusal2024secure_Benchmark_SECURE, liu2023secqa_Benchmark_SecQA, tihanyi2024_CyberMetric, liu2024_CyberBench}, SecBench also includes short-answer questions (SAQs) to present a more comprehensive evaluation.
This is because SAQs tend to be more challenging than MCQs: for MCQs, the LLM only needs to choose the correct answer(s) from the given options, while for SAQs, the LLM is prompted to construct its own answer based on the given question.
As a result, SAQs can evaluate the capability of the LLM at a higher level, especially considering the inherent limitations of LLMs (e.g., hallucinations and repetition).

\textbf{Multi-Domain.}
The questions in SecBench consist of 9 different domains, including \emph{D1. Security Management}, \emph{D2. Data Security}, \emph{D3. Network and Infrastructure Security}, \emph{D4. Security Standards and Regulations} , \emph{D5. Application Security}, \emph{D6. Identity and Access Control}, \emph{D7. Fundamental Software and Hardware and Technology}, \emph{D8. Endpoint and Host Security}, \emph{D9. Cloud Security}.
Particularly, the above domains were devised from several rounds of brainstorming and revision, which were expected to cover most (if not all) related sub-domains in cybersecurity.
Note that we do not expect these domains to be \emph{orthogonal}, and it is possible that one question can be reasonably labeled into different domains.
In our dataset, one question is assigned only one most-related domain label from D1 to D9.

\textbf{Example.}
For each line of data, it is either an MCQ or SAQ, provided with the question stem and corresponding answer, labeled with language (Chinese or English), level (Knowledge Retention or Logical Reasoning) and domain (from D1 to D9).

Following is one MCQ example, labeled in the domain of \emph{Security Management} and the level of \emph{Logical Reasoning}.
For MCQs, A blank is left in question stem, and there are four choices given in \emph{answers} for the tested LLM to select, with \emph{label} referring to the correct choice(s) among the four.

\small
\noindent\fbox{%
    \parbox{0.96\textwidth}{%

    \emph{\{"\textbf{question}":"In an information security risk management activity of a unit, the risk assessment report suggested that there were high-risk vulnerabilities in the FTP service of Server A. Subsequently, the unit chose the treatment measure of shutting down the FTP service in risk treatment, so may I ask to which type of risk treatment this measure belongs to ()","\textbf{answers}":["Risk reduction","Risk avoidance","Risk transfer","Risk acceptance"],"\textbf{label}":"B","\textbf{language}":"English","\textbf{domain}":"Security Management","\textbf{level}":"Logical Reasoning"\}}

    }%
}
\vspace{+0.3ex}
\normalsize

Following is one SAQ example, labeled in the domain of \emph{Data Security} and the level of \emph{Knowledge Retention}.
For SAQs, there is no choice given for selection, and the tested LLM is expected to construct the answer from scratch. in SAQ, \emph{answer} refers to the correct answer of the question stem, which will be used to evaluate LLM's output.

\small
\noindent\fbox{%
    \parbox{0.96\textwidth}{%

\emph{\{"\textbf{question}":"How does email encryption contribute to regulatory compliance and data protection efforts, and what are some common encryption methods used to secure email communications?","\textbf{answer}":"Email encryption helps organizations comply with data protection regulations such as GDPR, HIPAA, and CCPA by safeguarding sensitive information transmitted via email and preventing unauthorized access or disclosure. Common encryption methods include symmetric encryption, asymmetric encryption, and end-to-end encryption, each offering varying levels of security and usability.","\textbf{language}":"English","\textbf{domain}":"Data Security","\textbf{level}":"Knowledge Retention"\}}
    }%
}
\vspace{+0.3ex}
\normalsize

\begin{figure}[t]
\centering
\includegraphics[width = 0.99\linewidth]{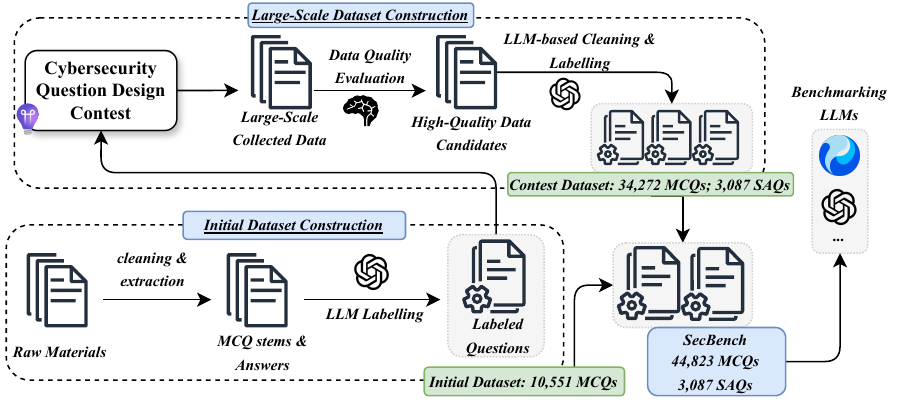}
\centering
\vspace{-2ex}
\caption{SecBench: Dataset Construction.}
\label{fig:secbench_dataset_construction}
\vspace{-1.6ex}
\end{figure}

\section{Dataset Construction Process}
\label{sec:dataset_process}

\subsection{Initial Dataset Construction}
\label{subsec:initial_dataset_contruction}

%
%


\textbf{Question Stems and Answers Extraction.}
We aim to construct a small batch of datasets to validate the rationality of the SecBench framework. To achieve this goal, we first collect raw materials from various sources that can be used to extract high-quality question data, including real exam questions from various cybersecurity fields, authoritative books, and so on.
Starting from these raw materials, we perform automated extraction of questions and answer data from these resources (for example, using regular expressions). After this step, we have collected a total of 10,551 high-quality MCQs, covering different domains.

\textbf{LLM-based Labeling.}
However, the dataset obtained in the previous step only contains question stems and answers, lacking the corresponding labels. Therefore, we used the powerful large language model - GPT4 \cite{gpt4}, to further label this part of the data.
With our carefully designed prompts, we utilized a powerful large model (GPT-4) to label these data, including tagging the difficulty level of the questions (whether it is Knowledge Retention or Logical Reasoning) and tagging the specific domain that the questions assess (as mentioned earlier, from D1 to D9).
After this step, we successfully labeled all collected questions. These questions became the prototype for SecBench, and we used this part of the labeled high-quality data to preliminarily validate the rationality of the SecBench design.

\subsection{Large-Scale Dataset Construction}
\label{subsec:large_dataset_contruction}

\textbf{Cybersecurity Question Design Contest.}
To further expand the SecBench dataset, we have organized a large-scale Cybersecurity Question Design Contest \cite{secbench_site}. In this contest, we expected participants to submit high-quality evaluation data across multiple domains, which we would subsequently clean and incorporate into the SecBench database.
Specifically, we categorized the data submitted in question into three quality levels, with different weight scores assigned to each level to encourage participants to submit high-quality data:

\emph{\textbf{ - Qualified Quality:}} The question meets the submission format, contains no factual errors, and is not duplicated with other questions submitted by the same or other participants.

\emph{\textbf{ - Medium Quality:}} The question has clear logic and expression, a well-defined domain, an unambiguous answer, and provides a clear and reasonable explanation along with a verifiable source.

\emph{\textbf{ - High Quality:}} The question design should have breadth, depth, and challenge, thus providing a high degree of differentiation for the capabilities of different models.


\textbf{Large-Scale Data Cleaning and Labeling.}
With the huge amount of data collected from the contest, we first manually assign the quality level (qualified, medium or high, as stated above) to each submission.
This process is performed by experienced experts with enough years of work experience in the cybersecurity domain, ensuring the justification of the quality attribution process.
%
%
Then, a rule-based filtering of these high-quality questions was performed to rule out possible duplications or incomplete data that were missed by the former human-based quality attribution.
Finally, similar to Sec.\ref{subsec:initial_dataset_contruction}, we labeled the questions by their level (KR or LR), language and domain with the help of LLM.
After the whole process, we obtained a total of 34,277 MCQs and 3,087 SAQs, greatly expanding our initial dataset quantitatively (more MCQs) and qualitatively (introducing the new evaluation form - SAQs).

\subsection{Dataset Distribution}
\label{subsec:dataset_distribution}

\begin{figure}[t]
\centering
\includegraphics[width = 0.99\linewidth]{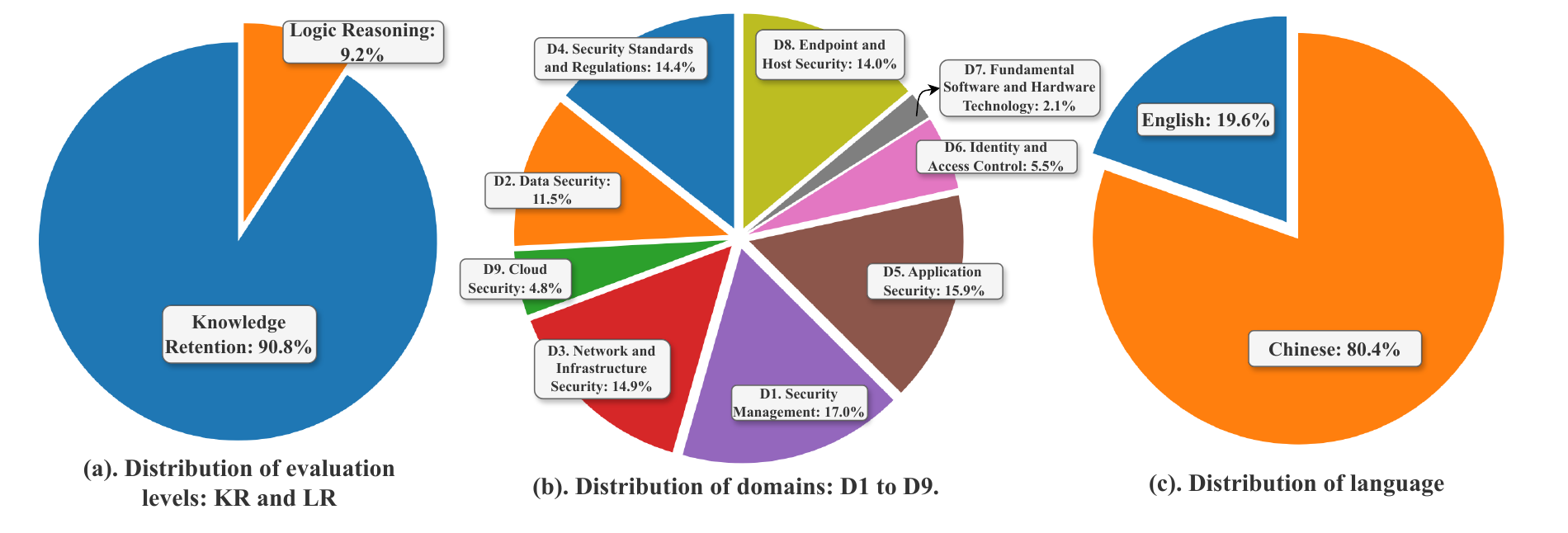}
\centering
\vspace{-3ex}
\caption{The distribution of evaluation level, domain and language of the 44,823 MCQs.}
\label{fig:distribution_MCQs}
\vspace{-2.6ex}
\end{figure}

\textbf{MCQ.}
Fig.\ref{fig:distribution_MCQs} shows the data distribution of the 44,823 MCQs in SecBench.
According to Fig.\ref{fig:distribution_MCQs}(a), the majority of the MCQs fall into the KR category (90.8\%), which is reasonable because MCQs tend to have short question stems and focus on testing the knowledge base of the LLM. Notably, 9.2\% of the MCQs are of the LR type, requiring the LLM to perform reasoning to obtain the correct answer.
As shown in Fig.\ref{fig:distribution_MCQs}(b), the data distribution across the 9 domains is generally even, with D6, D7, and D9 having relatively less data (5.5\%, 2.1\%, and 4.8\%, respectively). 
It is important to note that, given the large size of the dataset (44,823 MCQs), even a 4.8\% share corresponds to over 2,000 MCQs.
According to Fig.\ref{fig:distribution_MCQs}(c), the majority (80.4\%) of the MCQs are in Chinese, reflecting the context of the cybersecurity question design contest. 
Additionally, the 19.6\% of English MCQs (nearly 9,000 questions) provide a sufficient dataset for evaluating the LLM's cybersecurity capabilities in an English context.

\begin{figure}[t]
\centering
\includegraphics[width = 0.99\linewidth]{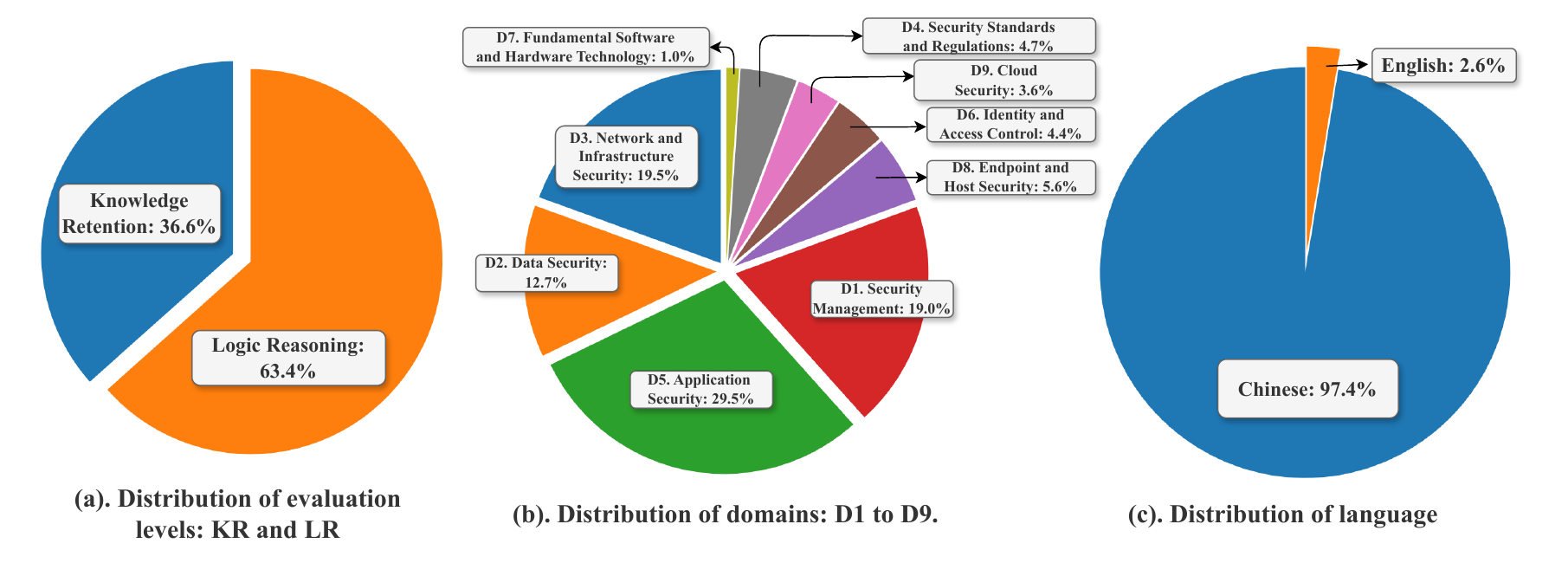}
\centering
\vspace{-3ex}
\caption{The distribution of evaluation level, domain and language of the 3,087 SAQs.}
\label{fig:distribution_SAQs}
\vspace{-2.6ex}
\end{figure}

\textbf{SAQ.}
Fig.\ref{fig:distribution_SAQs} shows the distribution of the evaluation level, domain, and language of the 3,087 SAQs.
As indicated in Fig.\ref{fig:distribution_SAQs}(a), 36.6\% of the SAQs are designed to assess knowledge retention, while 63.4\% are aimed at evaluating logic reasoning, indicating that the majority of SAQs are challenging and require the LLM to perform reasoning.
According to Fig.\ref{fig:distribution_SAQs}(b), the domains D1, D2, D3, and D5 constitute the majority of the assessed domains. Notably, even D7, which comprises only 1.0\% of the SAQs, includes 32 high-quality questions, given the overall dataset size of 3,087.
As shown in Fig.\ref{fig:distribution_SAQs}(c), 97.4\% of the SAQs are in Chinese, reflecting the context in which the contest was held, resulting in questions primarily constructed in Chinese.

\subsection{Benchmarking Process}
\label{subsec:dataset_benchmarking}

\textbf{MCQ.} 
The evaluation of MCQ is rather intuitive: for each MCQ, we check whether the model's output (i.e., model's choice(s) among A, B, C, and D) is the same as the correct answer.
For MCQs that involve more than one correct answer, model's output is judged as correct only when it is identical to the correct answer, meaning that no points are awarded for incorrect or incomplete selections.
Particularly, the evaluation of MCQ is implemented on the widely-used open-sourced evaluation framework - OpenCompass \cite{OpenCompass}.

\begin{figure}[t]
\centering
\includegraphics[width = 0.62\linewidth]{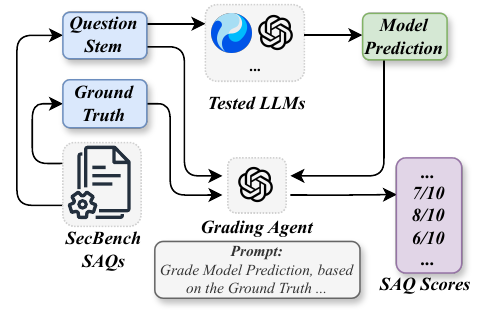}
\centering
\vspace{-3ex}
\caption{\textbf{SAQ evaluation process:} A sufficiently powerful LLM is used as the agent to grade the model prediction.}
\label{fig:SAQ_agent_process}
\vspace{-2ex}
\end{figure}

\textbf{SAQ.}
Grading an SAQ is not as intuitive as grading MCQ, in which case we only need to determine whether the LLM's choice is the correct one(s).
Particularly, grading SAQs requires to understand both the question stem and the model prediction (answer), and then fairly grade this prediction based on the ground truth, which is expected to huge amount of manual effort.
In our work, we introduce a \emph{Grading Agent} to realize the automatic grading of SAQs, and Fig.\ref{fig:SAQ_agent_process} shows the process of how the SAQs were evaluation on tested LLMs.
Specifically, each SAQ includes the question stem and the ground truth (i.e., correct answer) of the question.
The question stem is first fed into the tested LLMs to generate the \emph{Model Prediction}, which is the LLMs' answer waiting to be graded.
Then, the three parts of data, including the question stem, ground truth, and the model prediction will be given to the \emph{Grading Agent}, which is a sufficiently powerful LLM to grade the \emph{Model Prediction} based on the ground truth, and output the corresponding scores.
Specifically, this \emph{Grading Agent} should 1). be capable of fairly grading the model prediction, and 2). generate stable output (e.g., a final grading score for further processing)
In our work, we choose the GPT-4o mini \cite{gpt_4omini} as the grading agent, which achieves the balance between the performance (sufficient for the above two goals) and cost.

\section{Evaluation}
\label{sec:evaluation}

Based on SecBench, we conducted extensive benchmarking on 16 SOTA LLMs, including the GPT series and competitive open-source ones.

\subsection{MCQ Benchmarking}

Table \ref{tab:evaluation_MCQ} presents the benchmarking results for the 44,823 MCQs. The values in each cell represent the correctness percentage for the corresponding category.
Overall, the correctness of KR is significantly higher than that of LR, demonstrating the rationale behind our design (i.e., Logical Reasoning is more challenging than Knowledge Retention).
The performance of smaller LLMs (with fewer than 10 billion parameters) is predictably lower than that of larger LLMs (with more than 30 billion parameters)
Notably, the Tencent Hunyuan-Turbo model \cite{hunyuan} outperforms all existing models, including the state-of-the-art GPT-4o and GPT-4o-mini, achieving the highest correctness of 94.28\%. Its correctness in Logical Reasoning (93.06\%) is also significantly higher than that of other models, demonstrating Hunyuan-Turbo's strong capability in solving complex questions in cybersecurity.

\begin{table*}[t]
\centering
\caption{\textbf{MCQ Benchmarking:} The average correctness (values are percentage numbers) for all 44,823 MCQs. \textbf{Average:} Average correctness of all MCQs. \textbf{Level:} KR - Knowledge Retention; LR - Logical Reasoning. \textbf{Language:} CH - Chinese; EN - English. \textbf{Domain:} Sub-domains from D1 to D9 in Fig.\ref{fig:secbench_dimensions}.}
\label{tab:evaluation_MCQ}
\vspace{-1.6ex}
\scalebox{0.70}{
\begin{tabular}{lcccccccccccccc}
\toprule
& & \multicolumn{2}{c}{\textbf{Level}} & \multicolumn{2}{c}{\textbf{Language}} & \multicolumn{9}{c}{\textbf{Domain}}  \\
\cmidrule(lr){3-4} \cmidrule(lr){5-6} \cmidrule(lr){7-15}
 & \textbf{Average} & KR & LR & CH & EN & D1 & D2 & D3 & D4 & D5 & D6 & D7 & D8 & D9 \\
\midrule
o1-preview\cite{gpt_o1} & \textbf{89.52} & 89.92 & 85.86 & 90.77 & 85.96 & 88.37 & 89.08 & 89.46 & 87.85 & 91.99 & 88.28 & 91.23 & 91.33 & 87.95 \\
o1-mini\cite{gpt_o1-mini} & \textbf{90.50} & 90.92 & 86.28 & 92.56 & 82.00 & 88.77 & 89.78 & 90.02 & 90.39 & 93.55 & 88.56 & 88.54 & 92.36 & 87.66 \\
GPT-4o\cite{gpt_4o} & \textbf{90.99} & 91.82 & 82.75 & 92.87 & 83.23 & 90.32 & 90.32 & 88.71 & 90.51 & 94.13 & 89.61 & 84.43 & 93.54 & 90.00 \\
GPT-4o-mini\cite{gpt_4omini} & \textbf{88.79} & 89.86 & 78.27 & 91.37 & 78.17 & 88.07 & 88.30 & 84.71 & 88.86 & 92.84 & 86.91 & 75.78 & 92.90 & 87.33 \\
GPT-3.5-Turbo\cite{gpt_3.5_turbo} & \textbf{86.36} & 87.26 & 77.43 & 89.25 & 74.44 & 84.71 & 84.64 & 82.40 & 88.34 & 91.04 & 81.47 & 73.84 & 91.08 & 84.39 \\
\midrule
GLM-4-9B-Chat\cite{glm_4} & \textbf{84.57} & 85.14 & 78.95 & 88.26 & 69.38 & 83.26 & 81.41 & 80.18 & 87.11 & 89.01 & 80.43 & 71.35 & 89.82 & 83.23 \\
Llama-3-8B-Instruct\cite{llama3} & \textbf{77.71} & 78.43 & 70.58 & 80.70 & 65.43 & 77.26 & 74.34 & 73.07 & 77.48 & 82.85 & 74.51 & 62.92 & 84.14 & 76.74 \\
DeepSeek-V2-Lite\cite{deepseek_v2} & \textbf{79.07} & 80.07 & 69.22 & 83.40 & 61.26 & 78.21 & 74.86 & 73.73 & 82.48 & 84.56 & 71.85 & 65.51 & 85.23 & 76.60 \\
Qwen2-7B-Instruct\cite{Qwen2} & \textbf{87.74} & 88.29 & 82.29 & 90.77 & 75.29 & 86.94 & 85.79 & 85.20 & 89.38 & 90.69 & 83.41 & 82.49 & 91.86 & 83.74 \\
Yi-1.5-9B-Chat\cite{Yi_1.5} & \textbf{86.44} & 87.03 & 80.57 & 89.04 & 75.74 & 85.58 & 85.61 & 83.90 & 87.19 & 89.74 & 83.93 & 80.76 & 89.63 & 82.22 \\
\midrule
Hunyuan-Turbo\cite{hunyuan} & \underline{\textbf{94.28}} & \underline{94.41} & \underline{93.06} & 95.58 & 88.95 & 94.28 & 93.85 & 93.81 & 93.38 & 95.77 & 94.44 & 93.73 & 95.51 & 91.06 \\
Llama-3-70B-Instruct\cite{llama3} & \textbf{88.86} & 89.46 & 82.97 & 90.95 & 80.28 & 87.27 & 87.95 & 85.95 & 88.54 & 92.81 & 87.96 & 81.95 & 92.44 & 87.24 \\
Qwen2-72B-Instruct\cite{Qwen2} & \textbf{92.41} & 92.71 & 89.50 & 94.50 & 83.83 & 91.90 & 91.55 & 91.04 & 93.01 & 94.56 & 90.78 & 90.05 & 94.26 & 89.13 \\
Yi-1.5-34B-Chat\cite{Yi_1.5} & \textbf{89.59} & 90.04 & 85.19 & 91.48 & 81.82 & 89.14 & 88.71 & 88.47 & 90.00 & 92.20 & 88.44 & 87.14 & 91.38 & 84.15 \\
Mixtral-8x7B\cite{mixtral} & \textbf{86.08} & 86.78 & 79.19 & 88.58 & 75.76 & 85.05 & 84.70 & 81.52 & 87.30 & 91.04 & 83.13 & 75.35 & 89.75 & 84.39 \\
DeepSeek-V3\cite{deepseek_v3} & \textbf{92.79} & 93.18 & 88.92 & 94.52 & 85.66 & 91.70 & 91.53 & 92.00 & 93.20 & 95.23 & 91.58 & 90.92 & 94.66 & 89.50 \\
\bottomrule
\end{tabular}
}
\vspace{-1ex}
\end{table*}

\begin{table*}[t]
\centering
\caption{\textbf{SAQ Benchmarking:} The average scores graded by the grading agent (converted to a percentage scale). \textbf{Average:} Average correctness of all 3087 SAQs. \textbf{Level:} KR - Knowledge Retention; LR - Logical Reasoning. \textbf{Language:} CH - Chinese; EN - English. \textbf{Domain:} Sub-domains from D1 to D9 in Fig.\ref{fig:secbench_dimensions}.}
\label{tab:evaluation_SAQ}
\vspace{-1.6ex}
\scalebox{0.70}{
\begin{tabular}{lcccccccccccccc}
\toprule
& & \multicolumn{2}{c}{\textbf{Level}} & \multicolumn{2}{c}{\textbf{Language}} & \multicolumn{9}{c}{\textbf{Domain}}  \\
\cmidrule(lr){3-4} \cmidrule(lr){5-6} \cmidrule(lr){7-15}
 & \textbf{Average} & KR & LR & CH & EN & D1 & D2 &  D3 & D4 & D5 & D6 & D7 & D8 & D9 \\
\midrule
o1-preview\cite{gpt_o1} & \textbf{89.24} & 88.47 & 89.69 & 89.04 & 96.75 & 89.75 & 90.59 & 87.59 & 89.13 & 88.78 & 89.92 & 86.00 & 90.00 & 93.83 \\
o1-mini\cite{gpt_o1-mini} & \textbf{87.50} & 85.99 & 88.37 & 87.32 & 94.50 & 88.29 & 88.67 & 84.51 & 87.20 & 87.56 & 88.46 & 84.69 & 88.67 & 93.24 \\
GPT-4o\cite{gpt_4o} & \textbf{85.17} & 84.37 & 85.63 & 84.95 & 93.25 & 85.18 & 86.91 & 83.19 & 84.04 & 85.14 & 85.55 & 83.44 & 85.78 & 90.36 \\
GPT-4o-mini\cite{gpt_4omini} & \textbf{82.49} & 81.17 & 83.25 & 82.26 & 91.12 & 82.55 & 84.18 & 79.52 & 81.44 & 82.70 & 84.31 & 81.56 & 84.34 & 87.12 \\
GPT-3.5-Turbo\cite{gpt_3.5_turbo} & \textbf{74.78} & 75.54 & 74.34 & 74.49 & 85.50 & 74.32 & 76.15 & 72.83 & 72.95 & 75.09 & 76.72 & 77.50 & 74.22 & 80.45 \\
\midrule
GLM-4-9B-Chat\cite{glm_4} & \textbf{66.26} & 65.06 & 66.95 & 65.91 & 79.38 & 67.47 & 67.32 & 62.35 & 67.26 & 66.21 & 67.66 & 66.25 & 67.17 & 73.15 \\
Llama-3-8B-Instruct\cite{llama3} & \textbf{62.39} & 60.48 & 63.50 & 62.11 & 73.12 & 62.77 & 66.40 & 58.57 & 64.11 & 61.94 & 61.32 & 61.56 & 62.89 & 69.19 \\
DeepSeek-V2-Lite\cite{deepseek_v2} & \textbf{44.84} & 47.09 & 43.55 & 44.55 & 55.75 & 43.78 & 44.44 & 45.80 & 34.73 & 44.77 & 47.15 & 49.38 & 49.02 & 50.00 \\
Qwen2-7B-Instruct\cite{Qwen2} & \textbf{59.99} & 53.39 & 63.79 & 60.14 & 54.25 & 63.67 & 64.71 & 53.67 & 62.61 & 57.75 & 60.80 & 55.94 & 63.78 & 67.57 \\
Yi-1.5-9B-Chat\cite{Yi_1.5} & \textbf{65.24} & 64.98 & 65.39 & 65.01 & 73.88 & 65.86 & 67.04 & 62.61 & 63.77 & 64.47 & 66.57 & 66.88 & 66.36 & 74.23 \\
\midrule
Hunyuan-Turbo\cite{hunyuan} & \textbf{82.13} & 79.64 & 83.56 & 81.94 & 89.00 & 82.89 & 84.52 & 79.77 & 81.78 & 81.45 & 84.38 & 81.25 & 83.35 & 83.96 \\
Llama-3-70B-Instruct\cite{llama3} & \textbf{68.12} & 65.93 & 69.47 & 69.48 & 20.62 & 71.20 & 67.35 & 65.81 & 70.87 & 67.64 & 72.67 & 71.61 & 67.31 & 63.14 \\
Qwen2-72B-Instruct\cite{Qwen2} & \textbf{82.13} & 75.93 & 85.70 & 81.96 & 88.25 & 84.82 & 85.45 & 76.89 & 86.85 & 80.52 & 83.21 & 80.00 & 85.14 & 86.13 \\
Yi-1.5-34B-Chat\cite{Yi_1.5} & \textbf{75.03} & 67.82 & 79.18 & 74.75 & 85.38 & 79.25 & 78.27 & 68.05 & 76.71 & 73.71 & 77.37 & 62.50 & 79.25 & 81.80 \\
Mixtral-8x7B\cite{mixtral} & \textbf{74.78} & 72.42 & 76.15 & 74.38 & 89.88 & 75.06 & 78.44 & 71.01 & 73.42 & 74.37 & 77.23 & 74.38 & 76.65 & 80.18 \\
DeepSeek-V3\cite{deepseek_v3} & \textbf{83.71} & 82.47 & 84.43 & 83.54 & 90.25 & 84.03 & 84.90 & 82.03 & 83.56 & 83.32 & 84.96 & 80.94 & 84.91 & 87.84 \\
\bottomrule
\end{tabular}
}
\vspace{-1ex}
\end{table*}

\subsection{SAQ Benchmarking}

%
%

Table \ref{tab:evaluation_SAQ} presents the benchmarking results for the 3,087 short-answer questions (SAQs). The values in each cell represent the average score, graded by the grading agent, on a percentage scale for the corresponding category. Compared to MCQs, a larger gap is observed between different LLMs, indicating that solving SAQs is more challenging than MCQs. This is because the tested LLMs are required to generate their own answers rather than simply choosing from given options.
Specifically, for SAQs, the o1-preview \cite{gpt_o1} and o1-mini \cite{gpt_o1-mini} achieved the highest scores (89.24 and 87.50, respectively), due to their advanced capability on reasoning.
Note that Tencent Hunyuan-Turbo model \cite{hunyuan} also outperforms most existing models, achieving a high score of 82.13, which is competitive with the GPT-4o-mini \cite{gpt_4omini} (82.49) and DeepSeek-V3 \cite{deepseek_v3} (83.71).



\section{Discussion}
\label{sec:discussion}

%
%
%
%

\textbf{Rationale for Using LLMs in the Process.}
We utilized GPT-4o \cite{gpt_4o} for labeling data during the construction phase of SecBench, and GPT-4o-mini \cite{gpt_4omini} for grading SAQs in the benchmarking phase.
To ensure that these two LLMs are capable of performing the tasks, we explicitly checked their output.
Specifically, we randomly sampled a mini-batch from the output for manual verification and validated that (1) GPT-4o can successfully label the questions into the corresponding capability level and sub-domain, and (2) GPT-4o-mini can fairly grade the LLMs' answers based on the ground truth.

\textbf{Language Bias.}
SecBench exhibits a language bias towards Chinese (80.4\% in MCQs and 97.4\% in SAQs) because the majority of the data in SecBench comes from the Cybersecurity Question Design Contest, which is held in a Chinese context.
To ensure the originality and best maintain the original meaning of the questions, we currently do NOT translate the Chinese questions into English or vice versa.
As a result, the scale of SecBench could be further doubled via a translation process (e.g., via another powerful LLM), which we leave as future work.
Additionally, note that considering the large scale of SecBench, it still offers a sufficient amount of English questions (nearly 9,000 MCQs and 100 SAQs) for benchmarking.


%
%
%

\section{Conclusion}
\label{sec:conclusion}

We propose SecBench, a multi-dimensional benchmarking dataset specifically designed to evaluate LLMs in the cybersecurity domain. SecBench addresses the limitations of existing benchmarks by including questions in various formats (MCQs and SAQs), at different capability levels (Knowledge Retention and Logical Reasoning), in multiple languages (Chinese and English), and across various sub-domains.
The dataset was meticulously constructed by collecting high-quality data from open sources and organizing a Cybersecurity Question Design Contest, resulting in 44,823 MCQs and 3,087 SAQs.
To ensure the quality and consistency of the dataset, we employed GPT-4 for data labeling and GPT-4o-mini as a grading agent for the automatic evaluation of SAQs. Benchmarking results demonstrate the usability and comprehensiveness of SecBench, making it arguably the largest and most detailed benchmark dataset for LLMs in the field of cybersecurity. 
More information about SecBench can be found at our website \cite{secbench_site}, and the dataset can be accessed via the artifact link \cite{SecBench_artifact}.
%


\bibliographystyle{ACM-Reference-Format}

\bibliography{Ref}


\appendix

\section{Detailed Prompts}
\label{appendix:full_prompt}

\textbf{LLM-based labeling.} Following is the prompt text that is implemented on GPT-4 for labeling SecBench data.
In the prompt, we explicitly detailed the requirement and offered few-shot example to ensure the performance.

\footnotesize
\noindent\fbox{%
    \parbox{0.96\textwidth}{%

\# Task Description

I will upload a question related to information security, and now I need you to help me annotate these questions. I need you to annotate these questions from two dimensions: 1. Assessing ability: whether the question assesses basic Knowledge Retention or more challenging Logical Reasoning ability. 2. Assessing domain: which specific subfield the question belongs to. Next, I will elaborate on these two requirements:

1.Assessing ability. First, you will classify each question into one of the following two categories: (a). Knowledge Retention question: This type of question tests whether the test taker has the relevant background knowledge through straightforward descriptions. The answers to these questions can be obtained directly by querying the knowledge base and do not involve reasoning processes. (b). Logical Reasoning question: This type of question presents the test taker with a specific scenario and requires the test taker to reason or calculate within that scenario to arrive at the correct answer. Compared to Knowledge Retention questions, these questions are more challenging.

2.Assessing domain. Next, you will annotate the specific domain that each question assesses into one of the following 10 categories: (1). Identity and Access Control, (2). Cloud Security, (3). Endpoint and Host Security, (4). Security Standards and Regulations, (5). Data Security, (6). Security Management, (7). Network and Infrastructure Security, (8). Fundamental Software and Hardware Technology, (9). Application Security, (10). Others. Note that if you believe a question cannot be classified into any of the categories (1)~(9), then classify it as (10). Others.

Finally, you will provide the reason and basis for your classification of the question.

---

\# Input Introduction

The questions I upload consist of the following format:

\{"question":"Which of the following is directly related to database security?","answers":["Granularity of access control","Size of the database","Number of attributes in the relation table","Number of tuples in the relation table"],"label":"A"\}

Each line of the file includes 3 elements: question is the main body of the question, answers are the four provided options, and label is the correct answer.

---

\# Output Requirements

For each question I upload, you will annotate it according to my requirements and add the annotation results directly to the original data. The annotation results should be in Chinese. You will insert the annotation results after each piece of data, including:

 - "assessed ability": whether it is Knowledge Retention or Logical Reasoning.

 - "assessed domain": which domain the knowledge being assessed belongs to.

 - "reason for labeling": the reason for your annotation, including the reason for the ability annotation and the reason for the domain annotation, all need to be explained. This explanation must be detailed and explain why you believe the question assesses knowledge from a specific domain, not just give a meaningless reason.

Using the example question from the "Data Introduction" section, the annotated result should be as follows:

\{
  "question": "Which of the following is directly related to database security?",
  "answers": ["Granularity of access control", "Size of the database", "Number of attributes in the relation table", "Number of tuples in the relation table"],
  "label": "A",
  "assessed ability": "Knowledge Retention",
  "assessed domain": "Data Security",
  "reason for labeling": "This question directly tests specific basic knowledge related to database security and does not involve logical reasoning, so it is labeled as knowledge memory; it can be directly seen from the question stem that this question tests specific knowledge of database security and should be classified under 'Data Security'."
\}

---

    }%
}
\vspace{+0.3ex}
\normalsize

\footnotesize
\noindent\fbox{%
    \parbox{0.96\textwidth}{%
    
\# Judgment Criteria

For the first annotation task (i.e., 1. Assessing ability), you will follow the following criteria:

(1). Questions involving numerical calculations (e.g., encryption and decryption algorithms) must be Logical Reasoning questions.

(2). Questions involving specific code or Linux commands must be Logical Reasoning questions.

(3). If the question stem provides a hypothetical subject (e.g., specific names, a company, an organization, a security analyst) and describes this hypothetical subject in detail, and the question is designed based on this, then the question must be a Logical Reasoning question.

(4). Questions with relatively long stems are more likely to be Logical Reasoning questions (because the stem contains more information, often requiring the test taker to understand and reason). Conversely, questions with shorter stems are more likely to be Knowledge Retention questions. Note: This criterion is not absolute and can only serve as an auxiliary judgment criterion.

---

\# Few-shot Examples

\{
  "question": "Interface testing could involve which of the following?",
  "answers": ["The application programming interface (API)", "The graphical user interface (GUI)", "Both of the above", "None of the above"],
  "label": "C",
  "assessed ability": "Knowledge Retention",
  "assessed domain": "Application Security",
  "reason for labeling": "This question directly tests whether the candidate understands the specific process of interface testing and does not involve complex logical reasoning, so it is labeled as knowledge memory; interface testing is often related to specific applications, so it is classified under 'Application Security'."
\}

\{
  "question": "Two-key triple DES encryption: C=CK1[DK2[EK1[P]]], K1$\neq$K2, where the effective key is ()",
  "answers": ["56", "128", "168", "112"],
  "label": "D",
  "assessed ability": "Logical Reasoning",
  "assessed domain": "Data Security",
  "reason for labeling": "This question requires the candidate to understand DES encryption and calculate the correct answer based on the question stem, involving a logical reasoning process, so it is labeled as logical reasoning; the question tests encryption algorithms and is directly related to data security, so it is classified under 'Data Security'."
\}

...

---

Now please annotate the following question:

\{Input Question\}

    }%
}
\vspace{+0.8ex}
\normalsize

\textbf{SAQ Grading.} Following is the prompt that the grading agent (implemented on GPT-4o-mini in our work) used to grade the LLM's output for benchmarking SAQs.

\footnotesize
\noindent\fbox{%
    \parbox{0.96\textwidth}{%

Please help me grade a student's answers in the network information security exam. I will provide you with a dataset that contains three parts: 1. The question stem and specific question, 2. The standard answer (i.e., the full score answer), 3. The student's answer (to be graded). Specifically, you will perform the following steps:

1. For each question, read the question stem and understand the content of the question.

2. For each question, read the student's answer and compare it with the standard answer.

3. For each question, based on the differences between the student's answer and the standard answer, grade the student's answer. The question is scored on a 10-point scale, with a full score of 10 points.

Record and return the student's score for each question in the form of a JSON file.

---

Your output should only contain a JSON file in the following format, where each data entry only includes the student's score, with the key being "score" and the value being an integer between 0 and 10, inclusive, for example:

[\{"model\_score": 6\}]

---

Below are the questions you need to process, consisting of three parts: 1. The question stem and specific question, 2. The standard answer (i.e., the full score answer), 3. The student's answer (to be graded). Please grade based on the data below and return the JSON file in the format mentioned above:

1. The question stem:
\{Question Stem from SecBench\}

2. The standard answer (i.e., the full score answer):
\{Ground Truth from SecBench\}

3. The student's answer (to be graded):
\{LLM's output to be graded\}

    }%
}
\vspace{+0.3ex}
\normalsize

\end{document}